\documentclass[pra,twocolumn,showpacs,unsort,preprintnumbers,floatfix]{revtex4}
\usepackage{dcolumn,bm,amsmath,amssymb,esdiff}
\def\etal{\emph{et~al.}}

\newcommand{\Eref}[1]{Eq.~(\ref{#1})}
\newcommand{\tref}[1]{Table~\ref{#1}}
\newcommand{\rtw}{\rightarrow}
\newcommand{\cm}{\ensuremath{{\rm cm}^{-1}}}
\newcommand{\kms}{\ensuremath{k_{\rm MS}}}
\newcommand{\ksms}{\ensuremath{k_{\rm SMS}}}
\newcommand{\knms}{\ensuremath{k_{\rm NMS}}}
\newcommand{\fnm}[1]{\footnotemark[#1]}

\begin{document}

\author{V A Korol}
\email{vakorol@mail.ru}
\affiliation{Petersburg Nuclear Physics Institute, Gatchina, 188300, Russia}
\author{M G Kozlov}
\affiliation{Petersburg Nuclear Physics Institute, Gatchina, 188300, Russia}
\title{Relativistic corrections to isotope shift in light ions.}
\date{\today}


\begin{abstract}
We calculate isotope mass shift for several light ions using Dirac
wave functions and mass shift operator with relativistic corrections
of the order of $(\alpha Z)^2$. Calculated relativistic corrections
to the specific mass shift vary from a fraction of a percent for
Carbon, to 2\% for Magnesium. Relativistic corrections to the normal
mass shift are typically smaller. Interestingly, the final
relativistic mass shifts for the levels of one multiplet appear to
be even closer than for non-relativistic operator. That can be
important for the astrophysical search for possible
$\alpha$-variation, where isotope shift is a source of important
systematic error. Our calculations show that for levels of the same
multiplet this systematics is negligible and they can be used as
probes for $\alpha$-variation.
\end{abstract}

\maketitle

\section{Introduction}

Modern theories (such as the string theory and M-theory) predict
temporal and spatial variations of fundamental physical constants.
Some recent studies of quasar absorbtion spectra indicate that the
fine structure constant $\alpha=e^2/\hbar c$, where $e$ is the
charge of the electron and $\hbar$ and $c$ are the reduced Planck
constant and the speed of light respectively, could have changed
during the evolution of the Universe (see \cite{MFW03} and the
references therein). This result was not confirmed by other groups
\cite{VPI01,FVB03,QRL03,levshakov06a,Chand06}, and hence new
experiments are required. Laboratory experiments are rapidly
increasing their sensitivity to $\alpha$-variation (see
\cite{Pei06,Gin06} and references therein), but are currently
slightly less sensitive than astrophysical ones, where the time
scale is $10^{10}$ times longer. Finally, there may be some evidence
for varying $\alpha$ from the natural nuclear reactor in Oklo, which
operated about 2 billion years ago \cite{LT04}.

Astrophysical studies are based on the fact that atomic transition
frequencies depend on the parameter $x~=~(\alpha/\alpha_0)^2-1$, where
$\alpha_0 \approx 1/137$ is the laboratory value of $\alpha$.
Thus, one can look for $\alpha$-variation by comparing the frequencies
of atomic lines in the spectra of quasars with their laboratory values,
which correspond to $x=0$. One of the major systematic effects,
which can imitate $\alpha$-variation, is the isotope shift (IS). It
had been shown previously \cite{Lev94,MWF01,AMO03,FMG05} that
typical IS is of the same order of magnitude as the effect observed
in \cite{MFW03}.

In \cite{KKBD04}, a method to separate the effect of varying
$\alpha$ from the IS was proposed. This method is based on building
special combinations of atomic frequencies, some of which are
insensitive to both effects (``anchors''), while the other are
independent on the IS, but sensitive to $\alpha$-variations
(``probes''). One could then use the anchors to retrieve the actual
value of the redshift of the spectrum, and then use the probes to
evaluate the size of the variation $\Delta\alpha/\alpha$. In
addition, one can form the probes that are sensitive to the IS, but
not to the variation of $\alpha$. These probes could provide
information about isotopic abundances in the early Universe. Such
information should also be of interest to the astrophysicists,
because isotopic evolution is tightly linked with the general
evolution of the Universe.

Described method of anchors and probes requires precise calculations
of atomic relativistic coefficients $q$, which determine the
dependence of transition frequencies on $x$, and the IS coefficients
$k_\mathrm{IS}$. Factors $q$ strongly depend on the total angular
momentum $J$, which distinguishes levels of the same multiplet.
Relativistic corrections to the coefficients $k_\mathrm{IS}$ are
known to be small and their dependence on $J$ is much weaker.
Nevertheless in order to separate two effects reliably, one needs to
accurately account for relativistic corrections to IS. For light
atoms IS is dominated by the mass shift (MS). Usually MS
calculations are done with Dirac wave functions, but using
non-relativistic form of MS operator. In this paper we report
results of the MS calculations for 12 light atoms and ions.
We use the MS operator, which includes relativistic corrections of
the order of $(\alpha Z)^2$ \cite{SA94,ASY95,Sha98}. It is shown
that relativistic corrections to the operator partly compensate
corrections to the wave functions and final relativistic corrections
are smaller than in previous calculations. That may help to
distinguish between $\alpha$-variation and isotope effects in
astrophysics.

\section{Method}
\label{method}

\subsection{Theory of isotope shift}
\label{IS theory}
Total IS consists of MS and field shift (FS). The former accounts
for the finite mass of the nucleus, and the latter accounts for the
finite size of the nucleus. In the non-relativistic theory the MS is
further divided into normal mass shift (NMS), and specific mass
shift (SMS). The overall IS in the frequency of any atomic
transition of the isotope with mass number $A'$ with respect to the
isotope with mass number $A$ can be written as (see, for example
\cite{Kin84}):
\begin{equation}
\delta\nu^{A',A} =
(\knms + \ksms)
\left(
  \frac{1}{A'} - \frac{1}{A}
\right) +
F \, \delta\left\langle r^2\right\rangle^{A',A},
\label{IS}
\end{equation}
where $\left\langle r^2\right\rangle$ is the nuclear mean square
radius, $F$ is the FS coefficient and $\kms =\knms+\ksms$ is the MS
coefficient. In the non-relativistic theory NMS is described by
one-electron operator,
\begin{equation}
H_{\rm NMS}^{({\rm nr})} = \frac{1}{2M}\sum_i \bm{p}_i^2,
\label{NMS}
\end{equation}
where $\bm{p}_i$ is momentum of the $i$-th electron. Because of the
virial theorem $\knms$ has the simple form:
\begin{equation}
\knms^{({\rm nr})} = -\frac{\omega}{1823},
\label{KNMS}
\end{equation}
where the number 1823 refers to the ratio of the atomic mass unit
(amu) to the electron mass. SMS is described by a two-electron
operator:
\begin{equation}
H_{\rm SMS}^{({\rm nr})} = \frac{1}{M}\sum_{i<k} \bm{p}_i \cdot \bm{p}_k\,.
\label{Hsms}
\end{equation}

Consistent relativistic theory of MS can be formulated only within
quantum electrodynamics in a form of expansion in $\alpha Z$. It was
shown in Refs.~\cite{SA94,ASY95,Sha98} that the first two terms of
the expansion in $\alpha Z$ lead to the following relativistic MS
Hamiltonian:
\begin{equation}
H_{\rm MS}^{({\rm r})} =
  \frac{1}{2M}
  \sum_{i,j}
  \left\{
   \bm{p}_i \bm{p}_j -
   \frac{\alpha Z}{r_i}
   \left[
    \bm{\alpha}_i + (\bm{\alpha}_i\hat{\bm{r}}_i)\hat{\bm{r}}_i
   \right] \bm{p}_j
  \right\},
\label{HMSrel}
\end{equation}
where $\bm{\alpha}_i$ is the Dirac matrix of the $i$-th electron and
$\hat{\bm{r}}=\bm{r}/r$. As in the non-relativistic theory, one can
divide Hamiltonian (\ref{HMSrel}) into the one- and two-electron
terms:
\begin{align}
H_\mathrm{NMS} &= \frac{1}{2M} \sum_{i}
\left\{\bm{p}
- \frac{\alpha Z}{2r}
\left[\bm{\alpha}+
(\bm{\alpha}\cdot\hat{\bm{r}})\hat{\bm{r}}\right]\right\}_i^2\,,
\label{HNMSrel}\\
\begin{split}
H_\mathrm{SMS} &= \frac{1}{M} \sum_{i<k}
\left\{\bm{p}
- \frac{\alpha Z}{2r}
\left[\bm{\alpha}+
(\bm{\alpha}\cdot\hat{\bm{r}})\hat{\bm{r}}\right]\right\}_i
\\
&\times \left\{\bm{p}
- \frac{\alpha Z}{2r}
\left[\bm{\alpha}+
(\bm{\alpha}\cdot\hat{\bm{r}})\hat{\bm{r}}\right]\right\}_k\,.
\end{split}
\label{HSMSrel}
\end{align}
Non-relativistic expression \eqref{KNMS} is not applicable to
operator \eqref{HNMSrel} and now \knms\ has to be calculated on a
same footing as \ksms.

As pointed out above, for the light atoms and ions FS is much
smaller than MS. Still for the comparison with high precision
experiments one may need to account for FS. Below we focus on
calculating MS, but in the final \tref{final-is} we present FS
factors for some of the transitions.

\subsection{Electron correlations}
\label{CI+MBPT}

IS is very sensitive to electron correlations. High accuracy
calculations must account not only for correlations between valence
electrons, but also for core-valence correlations. Here we treat
both types of correlations within CI+MBPT method \cite{DFK96b}. In
this method the configuration interaction (CI) calculation is done
for valence electrons using effective Hamiltonian $H_\mathrm{eff}$,
which is formed within the second order many body perturbation
theory (MBPT) in the residual core-valence interaction. All our
calculations are done with modified Dirac-Fock code \cite{BDT77},
CI code \cite{KT87}, and MBPT code \cite{DFK96b}.

CI+MBPT method is easily reformulated for the IS calculations within
the finite-field approximation. In this approximation the IS
operator $H_{\rm IS}$ is added to the many-particle Hamiltonian $H$
with an arbitrary coefficient $\lambda$:
\begin{equation}
    H_\lambda = H + \lambda H_{\rm IS}.
\label{is1}
\end{equation}
The eigenvalue problem for Hamiltonian \eqref{is1} is solved
for $+\lambda$ and for $-\lambda$. Then, the IS correction to
the energy is recovered as:
\begin{equation}
    \Delta E_\mathrm{IS}=\frac{E_{+\lambda}-E_{-\lambda}}{2\lambda}.
\label{is2}
\end{equation}

All calculations are done independently for the field, normal, and
specific parts of the IS. Parameter $\lambda$ is chosen from the
considerations of the numerical stability and smallness of the the
non-linear terms. In the CI+MBPT calculations Hamiltonian
\eqref{is1} is used to construct $H_{\lambda,\mathrm{eff}}$. That
means that $H_\lambda$ is used on all stages of calculation starting
from solving Dirac-Fock equations for atomic core.

CI+MBPT method provides the solution of the eigenvalue problem for
valence electrons $H_{\rm eff} \Psi = E \Psi$, which is an
approximation to all-electron problem $H \Psi = E \Psi$. The latter
equation gives an exact solution for the given basis set.
Unfortunately this equation can be solved only for atoms with few
electrons and only for very short basis sets. Nevertheless, on a
short basis set we can compare all-electron and CI+MBPT calculations
with some simple calculation (for example, with one-configurational
approximation) and designate corresponding corrections as
$\Delta^\mathrm{short}$ and $\Delta_\mathrm{eff}^\mathrm{short}$.
Then we repeat CI+MBPT calculation on a much longer basis set and
find new correction $\Delta_\mathrm{eff}^\mathrm{long}$.
Although we can not make all-electron calculation for the long basis
set, we can use the short basis set calculation to shift the central
point of the final CI+MBPT calculation and estimate its error:
\begin{align}
\delta_\mathrm{shift} &\approx
  \Delta^\mathrm{short} - \Delta_\mathrm{eff}^\mathrm{short}\,,
\label{shift}\\
\delta_\mathrm{err} &\approx
  \delta_\mathrm{shift}
\frac{\Delta_\mathrm{eff}^\mathrm{long}}{\Delta_\mathrm{eff}^\mathrm{short}}\,.
\label{error}
\end{align}
We will use \Eref{shift} where possible to improve CI+MBPT results
and \Eref{error} to estimate theoretical error.

\section{Results of the calculations}
\label{calc-result}

\subsection{Lithium-like ions}
\label{li-like}
Lithium is the most simple atom in our consideration, since it has
only one core orbital $1s$ and one valence electron.
The simplicity of this system allows one to perform all-electron (full)
CI calculation for all three electrons.
Such calculation provides an exact solution on a
given one-electron basis set and allows to improve our results and
to control their accuracy using Eqs.~(\ref{shift},~\ref{error}). The
basis set consists of the orbitals: 1--9$s$, 2--9$p$, 3--8$d$, and
4--8$f$ (basis set $[9sp8df]$). CI+MBPT calculation is then done
with two longer basis sets: valence CI space is defined by the basis
set $[13sp12df]$ and all intermediate MBPT summations run over basis
set $[17sp18d19f20g21h]$ (see \cite{KP97} for details).

In \tref{Li-is} we compare our MS calculations with the experiment.
For Li the relative size of FS is about $10^{-5}$ and we neglect it.

\begin{table}[htb]
\caption { Isotope shifts (in MHz) in the spectrum of ~${}^6$Li,
with respect to ${}^7$Li. In the second column result of our CI+MBPT
calculation is given. Third column presents corrected values using
\Eref{shift} and the errors are estimated with the help of
\Eref{error}.}

\label{Li-is}

\begin{tabular}{cccl}
\hline
\hline
Transition &\multicolumn{2}{c}{Theory}& \multicolumn{1}{c}{Exper.}\\
           &CI+MBPT & Final \\
\hline $2s_{1/2}\rtw 2p_{1/2}$  & 10346& 10608(300) &
$10534.26(13)\footnotemark[1]$ \\
                         &           &       &
$10532.9(6)\footnotemark[2]$ \\
                         &           &       &
$10534.3(3)\footnotemark[3]$ \\
                         &           &       &
$10533.13(15)\footnotemark[4]$ \\
$2s_{1/2}\rtw 2p_{3/2}$  & 10346& 10607(300) &
$10533.3(5)\footnotemark[2]$ \\
                         &         &         &
$10539.9(1.2)\footnotemark[3]$ \\
                         &         &         &
$10534.93(15)\footnotemark[4]$ \\
$2s_{1/2}\rtw 3p_{1/2}$  & 14162& 14354(200) &
$14470(450)\footnotemark[5]$ \\
$2s_{1/2}\rtw 3p_{3/2}$  & 14162& 14398(250) &
$14470(450)\footnotemark[5]$ \\
$2s_{1/2}\rtw 3d_{3/2}$  & 13194& 13450(300) &
$13314(6)\footnotemark[6]$ \\
                          &        &         &
$13312(4)\footnotemark[7]$ \\
$2s_{1/2}\rtw 3d_{5/2}$  & 13194& 13450(300) &
$13314(6)\footnotemark[6]$ \\
                         &         &         &
$13312(4)\footnotemark[7]$ \\
$2s_{1/2}\rtw 4s_{1/2}$  & 14560& 14750(200) &
$14656(6)\footnotemark[6]$  \\
                         &         &         &
$14661(14)\footnotemark[7]$ \\
\hline \hline
\end{tabular}
\footnotetext[1]{Walls \etal \cite{WACLW03}}
\footnotetext[2]{Sansonetti \etal \cite{SRER95}}
\footnotetext[3]{Windholz and Umfer \cite{WU94}}
\footnotetext[4]{Scherf \etal \cite{SKJW96}}
\footnotetext[5]{Mariella \cite{Mar79}}
\footnotetext[6]{Lorenzen and Niemax \cite{LN82}}
\footnotetext[7]{Kowalski \etal
\cite{KNSWP78}}
\end{table}

We see that most experimental results for IS in Li are much more
accurate than our calculations. Corrections \eqref{shift} lead to
significant improvement of the agreement between calculations and
experiment. The error estimate \eqref{error} appears to be quite
reliable here as most calculated CI+MBPT values and all corrected
values agree with experimental data within the estimated errors. We
conclude that Eqs.~(\ref{shift},~\ref{error}) work nicely for
Lithium. On the other hand, even for such a simple system as Li, our
theoretical error for MS appears to be about 2\%, while the typical
error for the frequencies is only about 0.1\%. That shows the
extreme sensitivity of the MS to correlations and significantly more
important role of the high order MBPT corrections, which are
neglected here.

Calculations of other Li-like ions are done in a similar way using
basis sets of the same length. Corresponding results for
C~\textsc{iv}, N~\textsc{v} and O~\textsc{vi} are given in the final
\tref{final-is}.
There we also use Eq.~(\ref{error}) to estimate theoretical accuracy.

\subsection{B~II, C~III, and C~II}
\label{be-like}

We treat Be-like ions B~\textsc{ii} and C~\textsc{iii} as
two-electron ions with $1s^2$ core and the ground state
${}^1\!S_0\left[2s^2\right]$. Since the full 4-electron CI
calculation is more complex than in the case of 3-electron Li, a
shorter basis set $[5sp4df]$ is used here for such calculation. 
Valence 2-electron CI calculations were done with
$[10sp9df]$ and MBPT corrections were calculated using
$[16sp17d18f19g20h]$ basis set.

There are many experimental results for B~\textsc{ii}. \tref{BII-is}
presents the comparison of calculated and experimental IS for
B~\textsc{ii}. Unlike the case of lithium, here for most transitions
the estimated theoretical error is smaller than experimental one.
We see that generally there is good agreement between theory and
experiment.
Our calculation of IS in transition ${}^1\!P_1^o \rtw {}^1\!D_2$ is in good
agreement with the measurement \cite{LZJ98}, but is $3\sigma$ away
from the result of Ref.~\cite{vinti40}.
Note that for the transition ${}^1\!P^o_1 \rtw {}^1\!S_0$ the MBPT
correction for the long basis set appears to be anomalously large.
That leads to the much larger theoretical error, than for other
transitions.

\begin{table}[htb]
\caption{Isotope shifts (in MHz) for ${}^{11}$B with respect to ${}^{10}$B
      in the spectrum of B~\textsc{ii}.}

\label{BII-is}
\begin{tabular}{ccll}
\hline\hline
\multicolumn{2}{c}{Transition} &\multicolumn{2}{c}{IS}
\\
& $\lambda$ (\AA) & \multicolumn{1}{c}{Theory} &\multicolumn{1}{c}{Exper.}
\\
\hline
${}^1\!S_0[2s^2] \rtw {}^1\!P^o_1[2s2p]$ & 1362 & 21303(66)  & 21000(3000)\footnotemark[1] \\ & & & 21000(3000)\footnotemark[2]\\   
${}^3\!P^o_1[2s2p] \rtw {}^3\!P_2[2p^2]$ & 1624 & 22367(170) & 21600(3600)\footnotemark[1] \\ & & & 24000(3600)\footnotemark[2]\\   
${}^3\!P^o_0[2s2p] \rtw {}^3\!P_1[2p^2]$ & 1624 & 22369(170) & 21600(3600)\footnotemark[1] \\ & & & 21600(3600)\footnotemark[2]\\   
${}^3\!P^o_1[2s2p] \rtw {}^3\!P_1[2p^2]$ & 1624 & 22370(170) & 21600(6900)\footnotemark[1] \\ & & & 21600(6900)\footnotemark[2]\\   
${}^3\!P^o_2[2s2p] \rtw {}^2\!P_2[2p^2]$ & 1624 & 22369(170) & 21600(4500)\footnotemark[1] \\ & & & 18300(4500)\footnotemark[2]\\   
${}^3\!P^o_1[2s2p] \rtw {}^3\!P_0[2p^2]$ & 1624 & 22371(170) & 21600(3600)\footnotemark[1] \\ & & & 21600(3600)\footnotemark[2]\\   
${}^3\!P^o_2[2s2p] \rtw {}^3\!P_1[2p^2]$ & 1624 & 22372(170) & 21600(3600)\footnotemark[1] \\ & & & 21600(3600)\footnotemark[2]\\   
${}^1\!P^o_1[2s2p] \rtw {}^1\!S_0[2p^2]$ & 1843 & 12689(580) & 12600(1200)\footnotemark[2] \\   
${}^1\!P^o_1[2s2p] \rtw {}^1\!D_2[2p^2]$ & 3451 & 26663(55)  & 26600(100) \footnotemark[2] \\ & & & 26300(120)\footnotemark[3]\\   
\hline
\hline
\end{tabular}
\footnotetext[1]{J\"{o}nsson \etal \cite{JLZ98}}
\footnotetext[2]{Litzen \etal \cite{LZJ98}}
\footnotetext[3]{Vinti \cite{vinti40}}
\end{table}

C~\textsc{ii} is the only B-like ion which is interesting for
astrophysics. It is also of a particular interest to us because the
experiment \cite{Bur50} hints at approximately 2\% difference
between the shifts for the relativistic doublet. If this is
confirmed, it will mean that relativistic corrections to IS are
quite large even for light ions and should be included in accurate
calculations. That will also make isotope effects more dangerous for
the astrophysical searches for $\alpha$-variation.

The ground state of C~\textsc{ii} is
${}^2\!P^o_{1/2}\left[2s^22p_{1/2}\right]$. It has five electrons
and full CI becomes impractical even for the short basis set. That
makes it more difficult to determine theoretical error. Using our
calculations for Li-like and Be-like ions we estimate theoretical
error for C~\textsc{ii} to be about 1~--~2\%. On the other hand,
with relativistic MS operators \eqref{HNMSrel} and \eqref{HSMSrel}
the relative size of the shifts within one multiplet should be also
accurate to few percent. Our results shown in \tref{CII-is} indicate
that the difference between the shifts in this doublet must be
roughly two orders of magnitude smaller than hinted by the
experiment \cite{Bur50}. In fact, as we discuss below in more
detail, using relativistic operator \eqref{HMSrel} brings MS shifts
for the levels of one multiplet closer to each other than with
non-relativistic operators \eqref{NMS} and \eqref{Hsms}.

\begin{table}[htb]
\caption{ IS for ${}^{13}$C in respect to ${}^{12}$C in the doublet
line ${}^2\!S_{1/2}[2s2p^2] \rtw {}^2\!P_{J}^o[2s^23p]$ of
C~\textsc{ii}.
Here we use relativistic MS operators \eqref{HNMSrel} and
\eqref{HSMSrel}.
Calculation of \protect\citet{BFK06} was done with non-relativistic
operators. } \label{CII-is}
\begin{tabular}{rclcccc}
\hline
&&&&\multicolumn{3}{c}{IS (MHz)}
\\
\multicolumn{3}{c}{Transition}
& $\lambda$, \AA
& This work & Theory \cite{BFK06}
& Exper. \cite{Bur50}
\\
\hline ${}^2\!S_{1/2}$&$\rtw$&${}^2\!P_{3/2}^o$ &
2836.707 &$ -18185.0 $&$ -18500 $&$ -18350(60)$ \\   
${}^2\!S_{1/2}$&$\rtw$&${}^2\!P_{1/2}^o$ &
2837.605 &$ -18185.9 $&$ -18500 $&$ -18680(90)$ \\   
\hline
\hline
\end{tabular}
\end{table}

\subsection{Sodium-like ions and Mg I}
\label{na-like}
There are several Na-like ions which are important for astrophysics,
namely Na~\textsc{i}, Mg~\textsc{ii}, Al~\textsc{iii}, and
Si~\textsc{iv}. For Na~\textsc{i} and Mg~\textsc{ii} there are
experimental data available and we compare calculated MS with the
experiment in \tref{Na,Mg-is}.
Using nuclear radii from Ref.~\cite{Ang03} we have estimated FS for
$3s\rtw 3p$ transitions in ${}^{22-23}$Na~\textsc{i} and
${}^{24-26}$Mg~\textsc{ii} to be about $-2$ and $+16$ MHz,
respectively.
This is significantly smaller than the uncertainty in our MS
calculation.

In these calculations for Na~\textsc{i}, Al~\textsc{iii} and
Si~\textsc{iv} we use CI basis set $[14sp13df]$ and MBPT
basis set $[19sp18d19f20g21h]$. Our IS results for Si~\textsc{iv}
are within 1\% agreement with calculation of \citet{BDF03}, while
the differences for Mg~\textsc{i} are larger and can reach several
percent.

Neutral magnesium is one of the most well studied two-electron
atoms. The data from several experimental works on IS in
Mg~\textsc{i,ii} are presented in \tref{Na,Mg-is} and compared
with theoretical IS calculated on the basis sets $[12spd9f]$ and
$[17spdf9gh]$.

\begin{table}[htb]
\caption {Comparison with experiment for IS in Na~\textsc{i} and
Mg~\textsc{i,ii}. The FS is neglected (see text). All numbers are in
MHz.}

\label{Na,Mg-is}

\begin{tabular}{ccdr}
\hline
\hline
       Transition                 & $\lambda$, \AA &\multicolumn{1}{c}{MS}  & \multicolumn{1}{r}{IS (expt.)}\\
\hline
\multicolumn{3}{c}{${}^{22-23}$Na~\textsc{i}} \\
$     3s \rtw 3p_{1/2}          $ & 5896    &  775.8  & $758.5(7)\footnotemark[1]$ \\
                                  &         &         & $756.9(1.9)\footnotemark[2]$ \\
$     3s \rtw 3p_{3/2}          $ & 5890    &  776.5  & $757.72(24)\footnotemark[3]$ \\
\hline
\multicolumn{3}{c}{${}^{24-26}$Mg~\textsc{ii}} \\
$     3s \rtw 3p_{3/2}          $ & 2796    & 3086.3  & $3050(100)\footnotemark[4]$ \\
\hline
\multicolumn{3}{c}{${}^{24-26}$Mg~\textsc{i}} \\
${}^1\!S_0[3s^2]\rtw{}^3\!P^o_1[3s3p]$& 4571& 2686.9  & $2683.2(0)\footnotemark[5]$ \\
${}^1\!S_0[3s^2]\rtw{}^1\!P^o_1[3s3p]$& 2852& 1425.3  & $1415.3(5.0)\footnotemark[6]$ \\
${}^3\!P^o_0[3s3p]\rtw{}^3\!S_1[3s4s]$& 5167&  -367.9 & $-395.7(6.0)\footnotemark[7]$ \\
${}^3\!P^o_1[3s3p]\rtw{}^3\!S_1[3s4s]$& 5173&  -369.9 & $-391.2(4.5)\footnotemark[7]$ \\
${}^3\!P^o_2[3s3p]\rtw{}^3\!S_1[3s4s]$& 5184&  -373.1 & $-392.7(7.5)\footnotemark[7]$ \\
${}^3\!P^o_1[3s3p]\rtw{}^3\!D_2[3s3d]$& 3832&   58.7  & $60.6(3.0)\footnotemark[8]$ \\
${}^3\!P^o_2[3s3p]\rtw{}^3\!D_3[3s3d]$& 3838&   55.8  & $58.2(3.6)\footnotemark[8]$ \\
${}^1\!S_0[3s^2]\rtw{}^1\!P^o_1[3s4p]$& 2026& 2975.0  & $2918.9(1.2)\footnotemark[9]$ \\
\hline
\multicolumn{3}{c}{${}^{24-25}$Mg~\textsc{i}} \\
${}^1\!S_0[3s^2]\rtw{}^3\!P^o_1[3s3p]$& 4571& 1397.2  & $1405.2(0.1)\footnotemark[5]$ \\
${}^1\!S_0[3s^2]\rtw{}^1\!P^o_1[3s3p]$& 2852&  741.2  & $743.8(3.0)\footnotemark[6]$ \\
${}^3\!P^o_2[3s3p]\rtw{}^3\!S_1[3s4s]$& 5184&  -194.0 & $-203.9(7.5)\footnotemark[7]$ \\
${}^1\!S_0[3s^2]\rtw{}^1\!P^o_1[3s4p]$& 2026& 1547.0  & $1526.4(1.2)\footnotemark[9]$ \\
\hline
\hline

\footnotetext[1]{Pescht \etal \cite{PGM77}}
\footnotetext[2]{Huber \etal \cite{Huber78}}
\footnotetext[3]{Gangrsky \etal \cite{Gang98}}
\footnotetext[4]{Drullinger \etal \cite{DWB80}}
\footnotetext[5]{Sterr \etal \cite{SKM93}}
\footnotetext[6]{Beverini \etal \cite{Bev90}}
\footnotetext[7]{Hallstadius \& Hansen \cite{HH78}}
\footnotetext[8]{Hallstadius \cite{Hal79}}
\footnotetext[9]{Hanneman \etal \cite{Han06}}

\end{tabular}
\end{table}

\begin{table*}[p]
\caption{Isotope shift parameters for astrophysically important
transitions in Li-, Be-, B-, and Na-like ions, and in Mg~\textsc{i}.
The experimental values for the transition wavelengths were taken
from the NIST atomic spectra database (http://physics.nist.gov). The
SMS and NMS parameters are given in GHz$\cdot$amu, the FS parameter
$F$ is given in MHz/fm$^2$. Non-relativistic values for $\ksms$ and
$\ksms$ were obtained with operator \eqref{Hsms} and scaling
\eqref{KNMS} respectively.}

\label{final-is}

\begin{tabular}{rclrrrrdddc}
\hline
\hline

\multicolumn{3}{c}{Transition}
&$\lambda$, \AA
&\multicolumn{3}{c}{$\ksms$}
&\multicolumn{2}{c}{$\knms$}
&$F$
&$q$
\\
&&
&
&\multicolumn{1}{c}{non-rel.}
&\multicolumn{1}{c}{rel.}
&\multicolumn{1}{c}{other works}
&\multicolumn{1}{c}{non-rel.}
&\multicolumn{1}{c}{rel.}
&
&
\\
\hline
\multicolumn{11}{c}{C~\textsc{iv}} \\
$ 2s_{1/2}$&$\rtw$&$2p_{1/2}$ &~1551 & $-4509$
&$-4498(45)$&$-4511(23)\fnm{1}$
& -1060.5 &-1061.1   & -174 & 104\fnm{4} \\  
$ 2s_{1/2}$&$\rtw$&$2p_{3/2}$ & 1548 & $-4502$ &$-4499(45)$&$-4504(23)\fnm{1}$
& -1062.3 &-1061.2   & -173 & 232\fnm{4} \\  
\hline
\multicolumn{11}{c}{C~\textsc{iii}} \\
${}^1\!S_0[2s^2]$&$\rtw$&${}^3\!P^o_0[2s2p]$
                              &1910 &$-3461$ &$-3453(7)$&$-3473\fnm{1}$
& -861.2  &-860.3    &      & 74\fnm{1}  \\    
${}^1\!S_0[2s^2]$&$\rtw$&${}^3\!P^o_1[2s2p]$
                              & 1909 & $-3459$  &$-3454(7)$   &$-3472\fnm{1}$
& -861.6  &-860.2    &      & 108\fnm{1}  \\   
${}^1\!S_0[2s^2]$&$\rtw$&${}^3\!P^o_2[2s2p]$
                              & 1907 & $-3456$  &$-3454(7)$   &$-3468\fnm{1}$
& -862.6  &-860.1    &      & 178\fnm{1} \\    
$ {}^1\!S_0[2s^2]$&$\rtw$&${}^1\!P^o_1[2s2p]$
                              & 977  & $-2778$  &$-2774(80)$  &$-2790\fnm{1}$
& -1683.3 &-1677.3   &      & 165\fnm{1} \\    
\hline
\multicolumn{11}{c}{C~\textsc{ii}} \\
${}^2\!P^o_{1/2}[2s^22p]$&$\rtw$&${}^2\!S_{1/2}[2s2p^2]$
                              & 1036 &          &$-1295$      &$-1321\fnm{1}$
& -1586.9 &-1591.1   &      & 168\fnm{1} \\
${}^2\!P^o_{1/2}[2s^22p]$&$\rtw$&${}^2\!D_{3/2}[2s2p^2]$
                              & 1335 &          &$-2636$      &$-2671\fnm{1}$
& -1232.3 &-1230.3   &      & 178\fnm{1} \\
${}^2\!P^o_{1/2}[2s^22p]$&$\rtw$&${}^2\!D_{5/2}[2s2p^2]$
                              & 1334 &          &$-2637$  &$-2672\fnm{1}$
& -1232.3 &-1230.4   &      & 181\fnm{1} \\
${}^2\!P^o_{1/2}[2s^22p]$&$\rtw$&${}^4\!P_{1/2}[2s2p^2]$
                              & 2325 &          &$-2971$      &$-2960\fnm{1}$
& -707.2  &-700.8    &      & 132\fnm{1} \\
${}^2\!P^o_{1/2}[2s^22p]$&$\rtw$&${}^4\!P_{3/2}[2s2p^2]$
                              & 2324 &          &$-2972$      &$-2958\fnm{1}$
& -707.6  &-700.8    &      & 158\fnm{1} \\
${}^2\!P^o_{1/2}[2s^22p]$&$\rtw$&${}^4\!P_{5/2}[2s2p^2]$
                              & 2322 &          &$-2973$ &$-2956\fnm{1}$
& -708.1  &-700.6    &      & 202\fnm{1} \\
\hline
\multicolumn{11}{c}{N~\textsc{v}} \\
$ 2s_{1/2}$&$\rtw$&$2p_{1/2}$ & 1243 & $-7123$ &$-7099(70)$&
& -1323.3 &-1324.7   & -417 & 196\fnm{5}\\    
$ 2s_{1/2}$&$\rtw$&$2p_{3/2}$ & 1239 & $-7109$ &$-7103(70)$&
& -1327.6 &-1325.7   & -386 & 488\fnm{5}\\    
\hline
\multicolumn{11}{c}{O~\textsc{vi}} \\
$ 2s_{1/2}$&$\rtw$&$2p_{1/2}$ & 1038 & $-10312$&$-10266(100)$&
& -1585.0 &-1587.0   & -719 & 340\fnm{4}\\    
$ 2s_{1/2}$&$\rtw$&$2p_{3/2}$ & 1032 & $-10283$&$-10273(100)$&
& -1593.7 &-1590.5   & -719 & 872\fnm{4}\\    
\hline
\multicolumn{11}{c}{Na~\textsc{i}} \\
$ 3s_{1/2}$&$\rtw$&$3p_{1/2}$ & 5896 &         &$-116$&$-108(24)\fnm{2}$
& -278.9  &-278.3    & -33  & 45\fnm{4}\\     
          &      &            &      &         &    &$-97\fnm{3}$       &
          &          &      & \\
$ 3s_{1/2}$&$\rtw$&$3p_{3/2}$ & 5890 &         &$-116$&$-107(24)\fnm{2}$
& -279.2  &-278.5    & -33  & 63\fnm{4}\\     
          &      &            &      &         &    &$-97\fnm{3}$       &
    &      && \\
\hline
\multicolumn{11}{c}{Mg~\textsc{ii}} \\
$ 3s_{1/2}$&$\rtw$&$3p_{1/2}$ & 2803 & $-387$   &$-378$     &$-373(12)\fnm{2}$
& -586.6  &-586.7   & -127 & 120\fnm{5}\\
          &      &            &      &          &           &$-362\fnm{3}$
  &         &  &    & \\
$ 3s_{1/2}$&$\rtw$&$3p_{3/2}$ & 2796 & $-381$   &$-378$     &$-373(6)\fnm{2}$
& -588.1  &-587.3   & -127 & 211\fnm{5}\\
          &      &            &      &          &           &$-361\fnm{3}$
  &         &  &    &\\
\hline
\multicolumn{11}{c}{Mg~\textsc{i}} \\
${}^1\!S_{0}[3s^2]$&$\rtw$&${}^1\!P^o_{1}[3s4p]$
                              & 2026 & $-102$   & $-101$    &$-108\fnm{6}$
& -811.6  & -816.9   &  -175 & 87\fnm{6}\\
${}^1\!S_{0}[3s^2]$&$\rtw$&${}^1\!P^o_{1}[3s3p]$
                              & 2852 & $129$    & $130$     & 134\fnm{6}
& -576.5  & -575.3   &  -175 & 94\fnm{6}\\
\hline
\multicolumn{11}{c}{Al~\textsc{iii}} \\
$ 3s_{1/2}$&$\rtw$&$3p_{1/2}$ & 1863 &          &$-835$     &
& -882.9  &-882.6   & -254 & 216\fnm{5}\\      
$ 3s_{1/2}$&$\rtw$&$3p_{3/2}$ & 1855 &          &$-834$     &
& -886.7  &-884.6   & -254 & 464\fnm{5}\\      
\hline
\multicolumn{11}{c}{Si~\textsc{iv}}  \\
$ 3s_{1/2}$&$\rtw$&$3p_{1/2}$ & 1403 &          &$-1510$    &$-1535(11)\fnm{2}$
& -1172.4 &-1171.1   & -497 & 362\fnm{5}\\     
$ 3s_{1/2}$&$\rtw$&$3p_{3/2}$ & 1394 &          &$-1507$    &$-1505(7)\fnm{2}$
& -1180.0 &-1175.6   & -480 & 766\fnm{5}\\     
\hline
\hline
\footnotetext[1]{Berengut \etal \cite{BFK06}}
\footnotetext[2]{Berengut \etal \cite{BDF03}}
\footnotetext[3]{Safronova \& Johnson \cite{SJ01}}
\footnotetext[4]{Berengut \etal \cite{BDFM04}}
\footnotetext[5]{Dzuba \etal \cite{dfw99a}}
\footnotetext[6]{Berengut \etal \cite{BFK05a}}

\end{tabular}
\end{table*}

\section{Discussion}
\label{Discussion}
Our final results for MS in astrophysically important transitions
are listed in \tref{final-is}. Where available we also give FS
parameters and results of other calculations of SMS and relativistic
$q$-factors. The latter are known to be significantly different for
the lines of one multiplet because the splittings within the
multiplet are caused by relativistic spin-orbit interaction and
scales as $(\alpha Z)^2$. That makes these splittings natural
candidates as probes for $\alpha$-variation. However, $J$-dependence
of IS may introduce some systematic errors as isotope abundances in
the Universe may vary significantly. Here we use relativistic MS
operators and show that $J$-dependence for MS is smaller than one
might expect from the simple considerations and from experiment
\cite{Bur50}. That means that multiplet splittings are indeed good
probes for $\alpha$-variation.

In several cases we also made SMS calculations using
non-relativistic operator \eqref{Hsms} (see \tref{final-is}).
Comparison with our relativistic values indicate that relativistic
corrections grow from a fraction of a percent for Carbon ions to
about 1~--~2\% for Mg~\textsc{i,ii}. It is interesting though, that
relativistic MS values for one multiplet tend to be closer to each
other than non-relativistic values. The difference between final MS
shifts within one multiplet appear to be much smaller than
relativistic corrections themselves. It should be mentioned that
non-relativistic SMS presented in \tref{final-is} was obtained by
applying operator \eqref{Hsms} only to the upper components of the
Dirac functions. If this operator is applied to both components, the
result tends to be noticeably farther from the correct relativistic
value \cite{Tup03}.

It should be mentioned that relativistic calculations of NMS require
special care. Indeed, even in the non-relativistic limit calculated
NMS will correspond to the scaling \eqref{KNMS} for theoretical,
rather then experimental frequency. For light many-electron atoms
the difference between experimental and calculated frequencies can
be larger than actual relativistic effects. To account for that one
can multiply calculated relativistic NMS by the ratio
$\omega_\mathrm{exper}/\omega_\mathrm{theor}$. The values presented
in \tref{final-is} include this correction.

Using $q$-factors from \tref{final-is} we can also form number of
probes for isotope abundances, which are insensitive to
$\alpha$-variations. There are several such probes for Carbon ions.
The following probe for C~\textsc{iv}:
\begin{eqnarray}
\label{pis1}
P_1^{\rm IS}
&=& 0.69\omega\!\left(2s_{1/2}\rtw2p_{1/2}\right)_{\rm C\,\textsc{iv}} \\
&-& 0.31\omega\!\left(2s_{1/2}\rtw2p_{3/2}\right)_{\rm
C\,\textsc{iv}} = 24470.71 \textrm{~\cm},
\nonumber
\end{eqnarray}
has the MS parameter $\kms=2112$~GHz$\cdot$amu. 
More probes can be constructed from C~\textsc{iii} lines:
\begin{eqnarray}
\label{pis2}
P_2^{\rm IS}
&=& 0.71\omega\!\left({}^1\!S_0 \rtw {}^3\!P_0^o\right)_{\rm C\,\textsc{iii}} \\
&-& 0.29\omega\!\left({}^1\!S_0 \rtw {}^3\!P_2^o\right)_{\rm C\,\textsc{iii}}
= 21970.95 \textrm{~\cm},
\nonumber \\
\label{pis3}
P_3^{\rm IS}
&=& 0.59\omega\!\left({}^1\!S_0 \rtw {}^3\!P_0^o\right)_{\rm C\,\textsc{iii}} \\
&-& 0.41\omega\!\left({}^1\!S_0 \rtw {}^3\!P_1^o\right)_{\rm C\,\textsc{iii}}
= 9416.36 \textrm{~\cm},
\nonumber \\
\label{pis4}
P_4^{\rm IS}
&=& 0.69\omega\!\left({}^1\!S_0 \rtw {}^3\!P_0^o\right)_{\rm C\,\textsc{iii}} \\
&-& 0.31\omega\!\left({}^1\!S_0 \rtw {}^1\!P_1^o\right)_{\rm C\,\textsc{iii}}
= 4404.14 \textrm{~\cm},
\nonumber
\end{eqnarray}
These probes have $\kms=1810$ GHz$\cdot$amu, 
$776$ GHz$\cdot$amu 
and $1593$ GHz$\cdot$amu, 
respectively. Finally, one can form two probes for Nitrogen
and Oxygen:
\begin{eqnarray}
\label{pis5}
P_5^{\rm IS}
&=& 0.71\omega\!\left(2s_{1/2}\rtw2p_{1/2}\right)_{\rm N\,\textsc{v}} \\
&-& 0.29\omega\!\left(2s_{1/2}\rtw2p_{3/2}\right)_{\rm N\,\textsc{v}}
= 33719.52 \textrm{~\cm} \nonumber \\
&&\qquad (\kms=3537 \textrm{ GHz$\cdot$amu}), 
\nonumber \\
\label{pis6}
P_6^{\rm IS}
&=& 0.72\omega\!\left(2s_{1/2}\rtw2p_{1/2}\right)_{\rm O\,\textsc{vi}} \\
&-& 0.28\omega\!\left(2s_{1/2}\rtw2p_{3/2}\right)_{\rm O\,\textsc{vi}}
= 42255.9 \textrm{~\cm} \nonumber \\
&&\qquad (\kms=5213 \textrm{ GHz$\cdot$amu}). 
\nonumber
\end{eqnarray}

\section{Conclusions}
\label{conclusions}
We have used the method of the effective Hamiltonian to calculate IS
for astrophysically important transitions in ions with several
valence electrons. We used relativistic MS operators and found out
that relativistic corrections for light ions are not suppressed
numerically and are of the order of $(\alpha Z)^2$, i.e. 1\% for
$Z\approx 10$. On the other hand, the difference between MS values
for one multiplet appears to be at least one order of magnitude
smaller.

This work was supported by the Russian Foundation for Basic
Research, grant No. 05-02-16914. VK also acknowledges support from
St.Petersburg government, grant No. M06-2.4D-220 and the grants of the
Dynasty foundation and the Alferov fund.


\end{document}